\documentclass[fleqn,12pt,twoside]{article}
\usepackage{epsfig}
\usepackage{amsmath}
\usepackage{amssymb}
\usepackage{espcrc1}

\title{Experiments in randomly
  agitated granular assemblies close to the jamming transition} 

\author{G. Caballero\address[lmdh]{Laboratoire des Milieux D\'esordonn\'es et
        H\'et\'erog\`enes, Case 86, 4, place Jussieu,  
        75252~Paris~Cedex~05, France}\address[mex]{Departamento de F\'{i}sica,
        Facultad de Ciencias, Universidad Nacional Aut\'{o}noma de M\'{e}xico,
        04510 M\'{e}xico, Distrito Federal, M\'{e}xico.},
A. Lindner\addressmark[lmdh], G. Ovarlez\address[go]{Laboratoire des
        Mat\'eriaux et Structures du G\'enie Civile (LCPC-ENPC), Cit\'e
        Descartes, 2, all\'ee Kepler, 77420 Champs sur Marne, France}, 
G. Reydellet\addressmark[lmdh],  J. Lanuza\addressmark[lmdh] and
        E. Clement\addressmark[lmdh]}
      
\begin{document}

\maketitle

\begin{abstract}
We present here the preliminary results obtained for two experiments on
randomly 
agitated granular assemblies using a novel way of shaking. 
First we discuss the transport properties of a 2D model system undergoing
classical 
shaking that show the importance of large scale dynamics for this type of
agitation and offer a local view of the microscopic motions of a grain. 
We then develop a new way of vibrating the system allowing for random 
accelerations smaller than gravity. Using this method we study the evolution
of the free surface 
as well as results from a light scattering method for a 3D model system. 
The final aim of these experiments is to investigate the ideas of effective 
temperature on the one hand as a function of inherent states and on the other 
hand using fluctuation dissipation relations.
\end{abstract}

\section{Introduction}

\label{sec:intro}

Strikingly, systems as different as dense emulsions, colloidal pastes, 
foams or granular matter have many rheological properties in common
\cite{LiuNature}. 
All these systems can
flow like fluids when a sufficiently high external stress is applied but jam
into an amorphous rigid state below a critical yield stress. This jamming
transition is associated with a slowdown of the dynamics which led Liu et al. 
\cite{LiuNature} to propose an analogy between the process of jamming and
the glass transition for glass-forming liquids. Although the nature of this
jamming transition is still unclear experimentally \cite{DaCruz02}, several
attempts were made to adapt the concepts of equilibrium
thermodynamics to athermal systems out of equilibrium
\cite{Edwards,Kurchan,Ono,AnitaGen,danna,Coniglio}. For packings made of grains
with a size larger than a few microns, thermal fluctuations are too
small to allow a free
exploration of the phase space. The grains are trapped into metastable
configurations. The system can not evolve until external mechanical
perturbations like vibration \cite{NowakPRE} or shear \cite{Pouliquen} are
applied allowing the grains to overcome energy barriers and triggering
structural rearrangements. In this case, the free volume and the
configurations accessible for each grain are capital notions that were used 
to define the
new concept of ''effective temperature'' \cite{Edwards}. It was proposed
recently that this notion could account for the transport properties in the
vicinity of a jammed state through a fluctuation dissipation
theorem \cite{Kurchan}.\\
A. Fierro {\it et al.} \cite{naples} treat this question in analogy with supercooled
liquids. In this case we understand by
inherent states those 
that do not evolve with time and that correspond to the local minima of the
potential energy in the 3N-dimensional configuration space of the particle
coordinates. A. Fierro et al. \cite{naples} now consider the
mechanically stable states of granular materials at rest as inherent
states. They work numerically with a 3D system of hard spheres subject to
gravity and undergoing a Monte-Carlo shaking. During the dynamics, the
system cyclically evolves for a time $\tau _{0}$ (corresponding to the tap
duration) 
at a
finite value of the bath temperature $T_{\Gamma }$ (corresponding to the tap
amplitude) and
is suddenly frozen at zero temperature in one of its inherent states. They
find that the system reaches a stationary state determined by the tap
dynamics (i.e. different $T_{\Gamma }$ and $\tau _{0}$). These stationary
states are indeed characterized by a \textit{single} thermodynamical
parameter since one finds a single master function when
the fluctuation $\overline{\Delta E^{2}}$ is plotted as a function of $%
\overline{E}$, where $E$ is the potential energy of the ensemble of grains.
Based on this result, they conclude that the quasi-stationary state can be
genuinely considered a``thermodynamical state'' and they define an effective
temperature through the fluctuation-dissipation relation.\\
 Here we design an
experimental set-up with the final aim to test closely the ideas of
effective temperature on the one hand as a statistics of inherent states and 
on the other hand as a result
of the fluctuation-dissipation relation by studying the transport properties
of a grain. Contrarily to previous work on vibrated granular assemblies, we design a new
way of shaking the granular material at accelerations much lower than 
gravity. In the first part of the paper, we study a model
granular assembly in 2D and use this preliminary investigation to design the
final 3D experiment. The preliminary results obtained with this set-up are presented in the
subsequent section.

\section{Vibration of a 2D model granular assembly}

\label{sec:2d}

\subsection{Experimental Set-Up}

\label{sec:2dsetup}

\begin{figure}[hh]
\begin{center}
\epsfxsize0.8\linewidth
\epsfbox{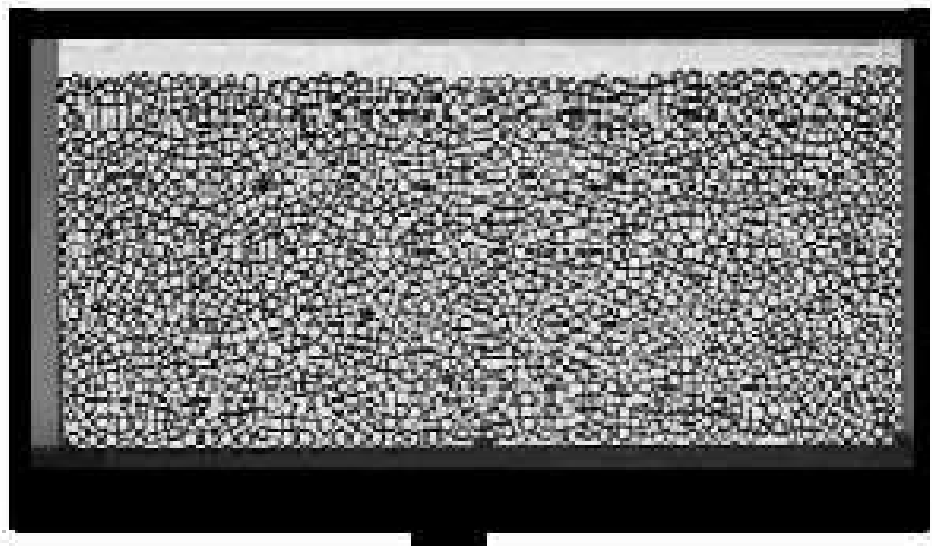}
\end{center}
\caption{Experimental set-up showing the 2D model system for a 
granular assembly under vibration: a layer of polydisperse particles is
confined between two vertical glass plates. A single sinusoidal tap is applied
every 5 seconds via an electromagnetic shaker.}
\label{fig:setup2d}
\end{figure}

We study the displacement of tracer particles in a 2D model granular assembly
which is exposed to tapping. To do so we use the following model system
(figure \ref{fig:setup2d}): a layer of polydisperse particles is confined
between two vertical glass plates. The dimensions of the glass plates are
20~cm times 30~cm. The lateral walls are made of Teflon. 
We use a mixture of small cylindrical steel particles of 3~mm height with
three different diameters, notably $d_{1}=6$~mm, $d_{2}=5$~mm and
$d_{3}=4$~mm to create a disordered packing. The cell is partially filled
which leads to about $30$ particles in the vertical direction and to $60$
particles in the horizontal direction.

We now study the response to tapping of the system by using an electromagnetic
shaker. A single sinusoidal tap is applied every 5s. The subsequent motion of
the shaker and the beads lasts about 1~s. Note that we can vary the
intensity of the tapping by changing the applied voltage from 100~mV to
500~mV leading to accelerations from approximately 
$\gamma_{peak}/g=1$ to $\gamma_{peak}/g=2$, with
$\gamma_{peak}$ being the peak acceleration. The system contains between $1$
to $3$ tracer particles of diameter $d_1=$~5~mm that can be 
positioned at a given initial position. A
CCD camera coupled to a computer captures a picture after a given number of
taps (typically between 500 to 1000 taps) and allows us to follow the long
time displacements of the tracers. We then extract the trajectories of the
tracer particles using image processing and finally obtain their $x$ and $y$
coordinates as a function of time.

Note that we typically observe a compaction of the initial surface 
occupied by the
grains of about 1$\%$ during the first 10000 taps, that correspond to the very
beginning of our experiments. Afterwards we observe fluctuations of the height
of the granular layer but can not detect further compaction within our
experimental resolution. 
 
\subsection{The large scale and long time convection dynamics}

\label{sec:2d large scale}

In the following, we describe the experimental results obtained when
performing experiments with a free surface at the top of the granular
assembly. In this case, for accelerations larger than the acceleration of
gravity and for a given phase of the motion, the grains on the upper surface
are launched freely with an upwards velocity given by the acceleration of the
cell. A bit later an impact with the rest of the granular assembly occurs as 
the latter catches up with the grains at the upper surface \cite{ClementRev}. In the
absence of boundaries all grain would follow about the same free flight
trajectories. This holds only if the grains come to a complete rest between
two taps and thus provided that the time for energy dissipation is 
smaller than the time between two impacts. 
The sequence of impacts is at the origin of the more or less
random shaking of the granular assembly and subsequently, is triggering the
compaction/decompaction phenomenology.

The problem is that this shaking procedure is strongly complicated by the
presence of frictional boundaries which are in our case the lateral walls made
of Teflon. The motion of the grains in contact with
the boundaries is perturbed by friction and for vertical boundaries 
these grains hit the rest of the assembly slightly before their
neighbors positioned further away from the boundaries. This leads to a slow but
inexorable descent of the grains at the boundaries since these grains are
likely to occupy the empty space left by their neighbors still in free
flight. Because of mass conservation, the compound of this motion, tap after
tap, leads to large scale convection rolls \cite{Evesque,Laroche,Clement92}.

Furthermore, the magnitude of the convection rolls depends not only on
grain/boundary friction values \cite{Clement92} but also on the packing density
values. The reason is that the higher the packing fraction, the more
efficient is the transmission of vertical forces to horizontal forces (given
by the
Janssen's effective parameter : see \cite{Ovarlez}). This effect, in
association with the friction coefficient, fixes a limit\ between an upper
part where the convection rolls are located and a bottom part which is blocked
since the boundary grains cannot overcome the Coulomb threshold. It was show
by a simple argument that this effect of a jammed phase localization depends
on the aspect ratio of the cell, it is reduced in the limit of small friction and
is bound to disappear for maximal accelerations larger than $2g$ \cite{Duran}.

\begin{figure}[h]
\begin{minipage}[b]{0.32\linewidth}
\begin{center}
\epsfxsize=\linewidth
\epsfbox{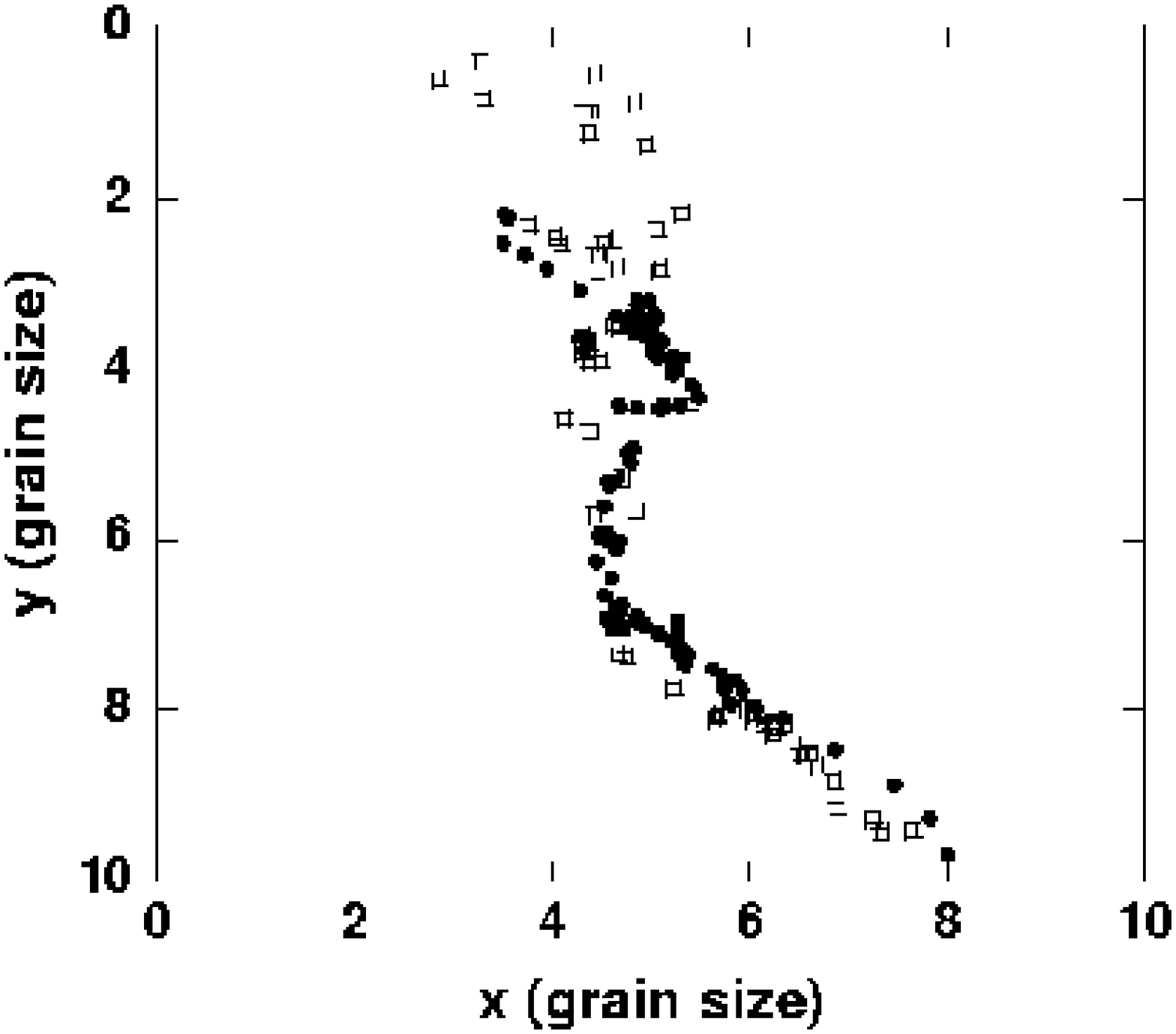}
\end{center}
\end{minipage}
\hfill\begin{minipage}[b]{0.32\linewidth}
\begin{center}
\epsfxsize=\linewidth
\epsfbox{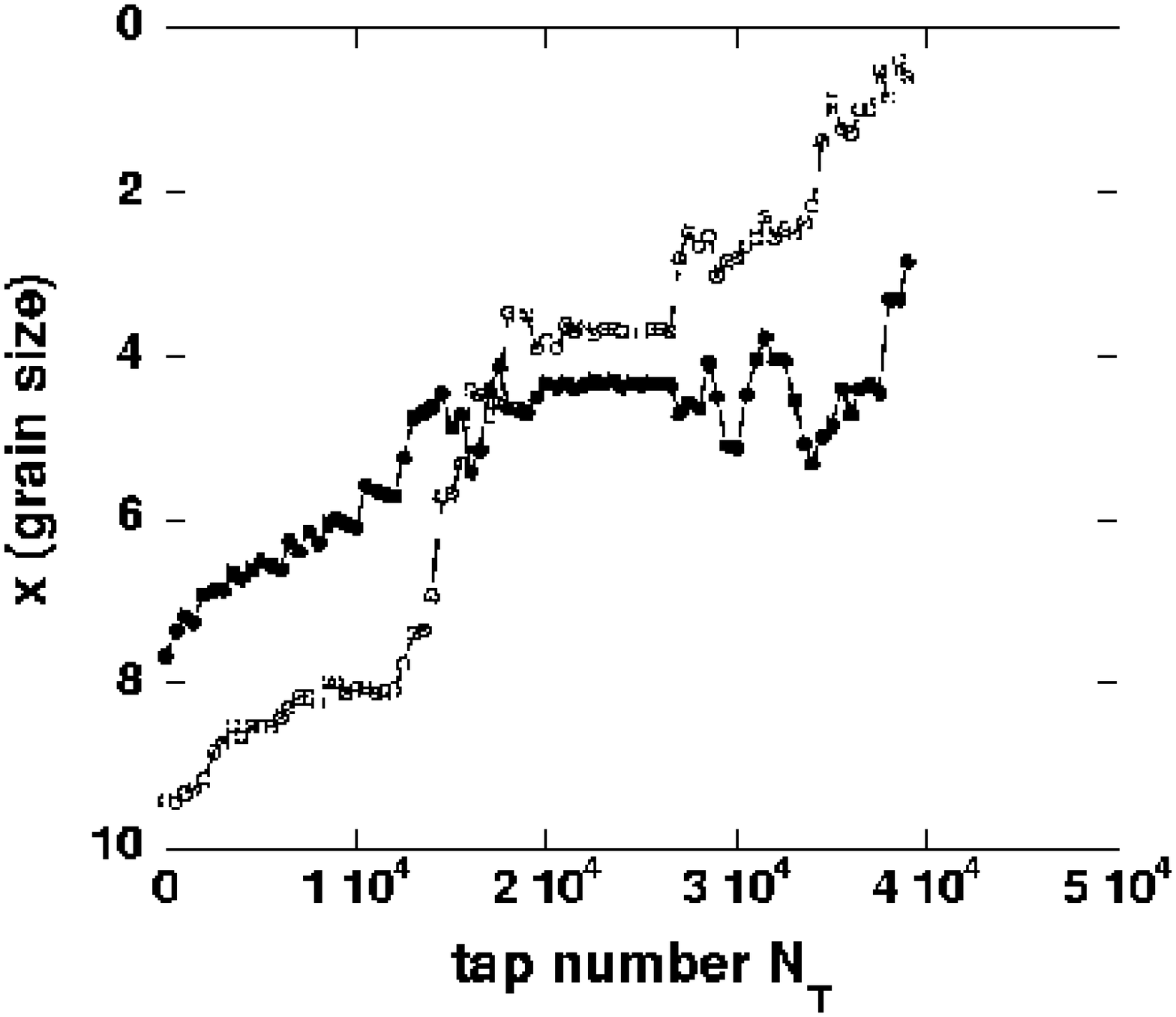}
\end{center}
\end{minipage}
\hfill\begin{minipage}[b]{0.32\linewidth}
\begin{center}
\epsfxsize=\linewidth
\epsfbox{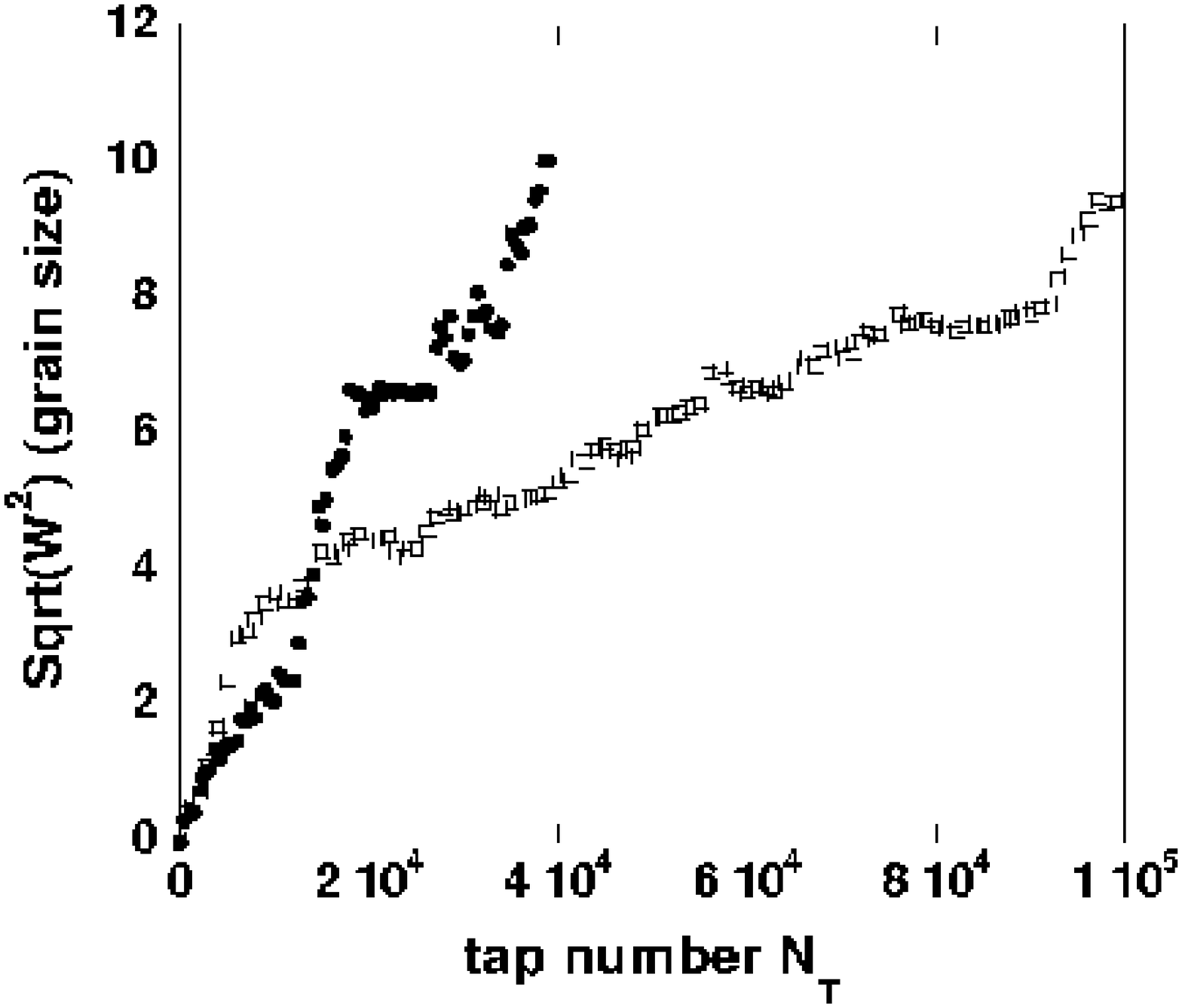}
\end{center}
\end{minipage}
\caption{Tracer displacement for two experiments having the same initial
 configuration but different tapping amplitudes, given by the applied voltage
  of 150~mV for experiment 1 and 325~mV for experiment 2. 
For experiment 1 a snapshot is taken every 1000 taps and for
  experiment 2 every 500 taps. 
All displacements are scaled on a typical grain size
  of 5~mm. The observed tracer is placed in the middle of the celle at the
  beginning of the experiments. Left: Trajectory of experiment 1 ($\square$) and experiment 2 
($\bullet$). Middle: $x$ ($\bullet$) and $y$ ($\circ$) coordinate 
for experiment 2
  as a function of the tap number. Right: $\sqrt{ \langle W^2 \rangle}$ 
as a function of tap
  number for experiment 1 ($\square$) and experiment 2 
($\bullet$) }
\label{fig:convection}
\end{figure}

First, we describe two experiments having the same initial
configuration but different tapping amplitudes. The steel grains are chosen to
have a low friction with the Teflon boundaries such that for all the experiments we
show, we do not observe a phase where the convective motion is blocked in the
bottom part. The experiment with the lower tapping amplitude is experiment 1
whereas the experiment at a higher tapping amplitude is experiment 2. The
left graph of figure \ref{fig:convection} shows the trajectories of the tracer
for the two experiments. The displacement is scaled on an average grain size
of 5~$mm$. One observes that the tracers follow in both cases
the same trajectory, signature of a convective motion without significant
diffusion of the particles. 
Note that for experiment 1 a snapshot (corresponding to
one data point) is taken every 1000 taps whereas for experiment 2 a snapshot is
taken every 500 taps. The velocity of the tracer is thus different from one
experiment to another as one would expect. Even if the two trajectories seem
quite smooth, the displacement of the tracer as a function of time is very
irregular. This is illustrated on the graph in the middle of figure
\ref{fig:convection} that shows the $x$ and $y$ coordinates of experiment 2. We
have observed for several experiments, that the tracer does nearly not move
for a large number of taps and then abruptly continues its displacement. The
graph on the right side of figure \ref{fig:convection} finally shows 
$\sqrt{\langle W^2 \rangle}$ the sliding average of the root
mean square displacement for experiments 1 and 2. This graph shows clearly
that the displacement as a function of the number of taps is less important
for experiment 1 than for experiment 2. Furthermore, 
it becomes again clear from this graph that large scale 
convective displacements are
taking place in the granular assembly. An estimation of the average particle
velocity observed in our experiments leads to values roughly between 
$v_{particle}\sim$~1 particle size per 1000
taps or even per 10.000 taps depending on the applied voltage. 
Interestingly, for grains of
typically the same size, Philippe et al. \cite{PhilippePRL} 
find the same order of
magnitude for their 3D tapping experiment. 
Note, that the convection velocity increases
with the tapping amplitude.

\begin{figure}[hh]
\begin{minipage}[b]{0.5\linewidth}
\begin{center}
\epsfxsize=\linewidth
\epsfbox{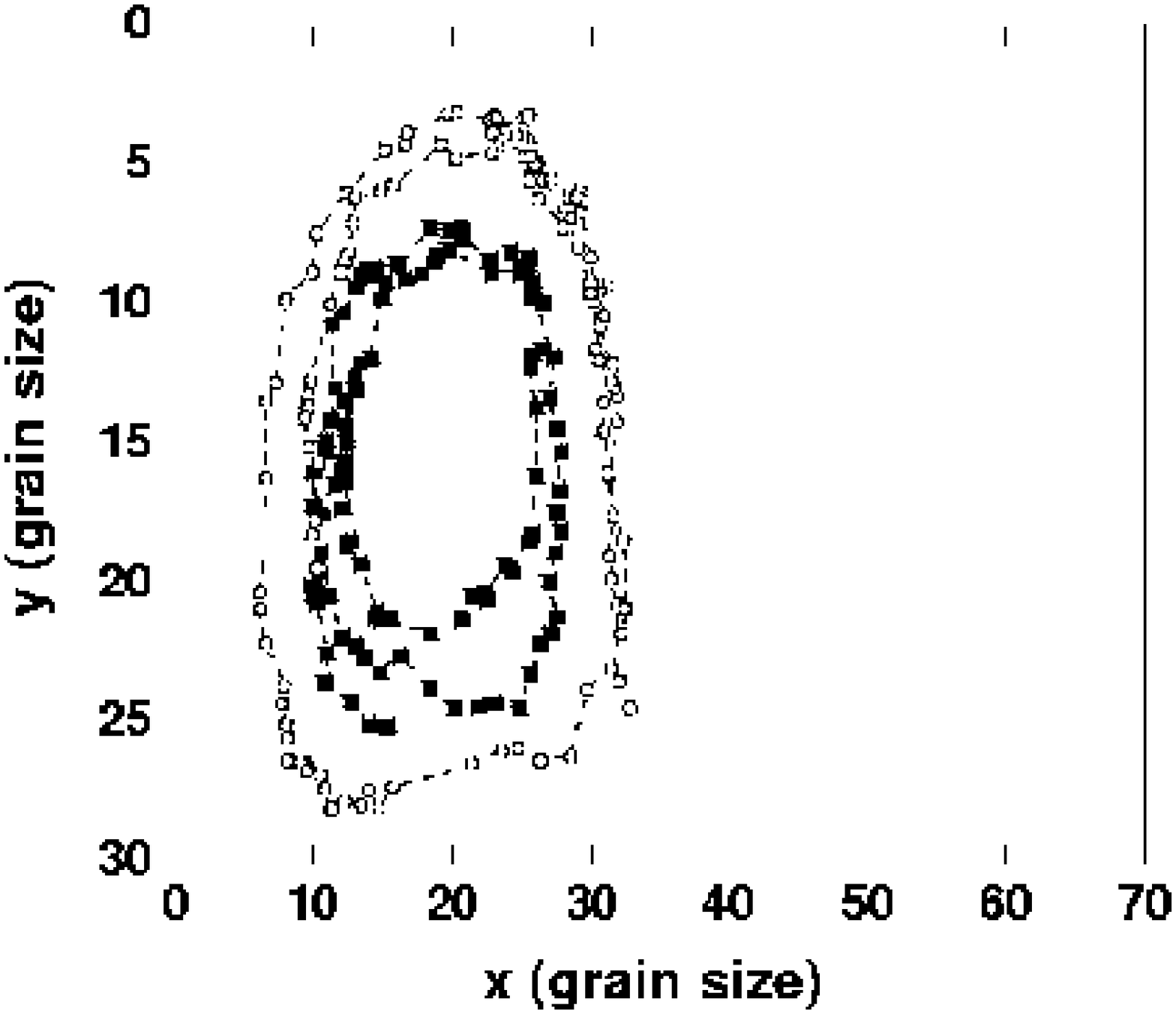}
\end{center}
\end{minipage}
\hfill\begin{minipage}[b]{0.5\linewidth}
\begin{center}
\epsfxsize=\linewidth
\epsfbox{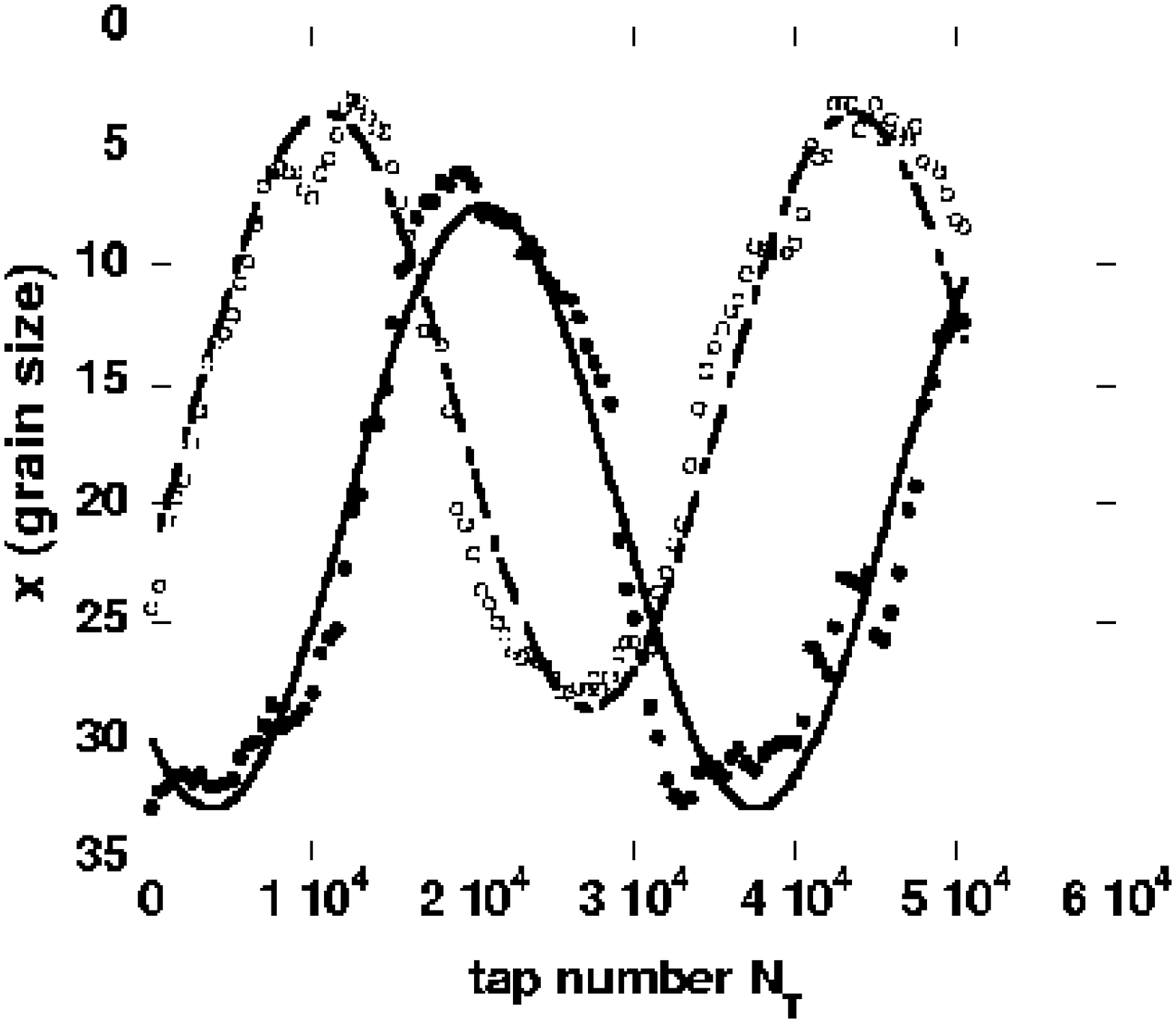}
\end{center}
\end{minipage}
\caption{Tracer displacement for experiment 3 with a high tapping amplitude
  (500~mV). A snapshot is taken every 500 taps. Left: Trajectories of tracer 
1 ($\blacksquare$) and tracer 2 ($\circ$) showing huge convection
rolls. Right: 
$x$ and $y$ coordinate of
  tracer 2 as a function of $N_T$.}
\label{fig:rolls}
\end{figure}

The importance of the large scale displacements becomes even more
clear when looking at a third experiment (experiment 3). In this case, the
tapping amplitude was increased further and three tracer particles were
followed. A snapshot was taken every 500 taps. On the left of figure \ref{fig:rolls} one
can see the trajectories of two of the three particles and one can conclude
that huge convection rolls form in the system. Note that the convection roll
observed occupies half the size of the cell. After one period the
particle follows nearly the same trajectory again. This is illustrated on the
right graph of figure \ref{fig:rolls} that shows the $x$ and $y$ coordinates
for one of the particles as a function of the number of taps. One may notice
that they follow nearly a perfect sinusoidal motion over more than one
period. Once again one can conclude that the motion is purely convective. 

Consequently, it is clear from these model experiments that the leading
dynamical behavior for the shaken grains is convection with grains following
well defined trajectories. Even if for low shaking
amplitudes we have intermittent dynamics as seen on figure
\ref{fig:convection} (middle) we did not observe any 
significant diffusive motion of the grains. In other words, the grains are
likely to keep their neighbors for a very long time. Thus, any attempt to
characterize the compaction dynamics by self diffusive properties of the
grains, is doomed to fail. Another important question that one could rise, is
whether the observed convection dynamics is related to the steady states
reported experimentally in tapping experiments 
\cite{NowakPRE,PhilippePRL,PhilippePRE,KnightPRE,KnightPRE96}. 
One might suggest that the
convection rolls lead to a decompaction of the system and that there is
thus a competition between this decompaction and compaction due to
shaking. This could explain the different results reported by Nowak and Knight
{\it et al.} \cite{NowakPRE,KnightPRE,KnightPRE96} and
Philippe et al. \cite{PhilippePRL,PhilippePRE} 
since they use columns with different aspect ratios.
Moreover, the column by the Chicago group is very narrow (about 15 grains
across) and the role of wall friction and finite size effects could then be
crucially important. All these questions are to us still open.

\subsection{Suppression of convection}
\label{sec:sup convec}

We now try to suppress the convection by putting a lid on top of
the cell at a distance of less than one bead diameter. The idea is to produce
a bounce of the grains on the lid almost immediately after the launch of the
packing. This bounce downwards is likely to suppress the time lag effect in
the grain trajectories due to the presence of walls with friction.

\begin{figure}[hh]
\begin{minipage}[b]{0.5\linewidth}
\begin{center}
\epsfxsize=\linewidth
\epsfbox{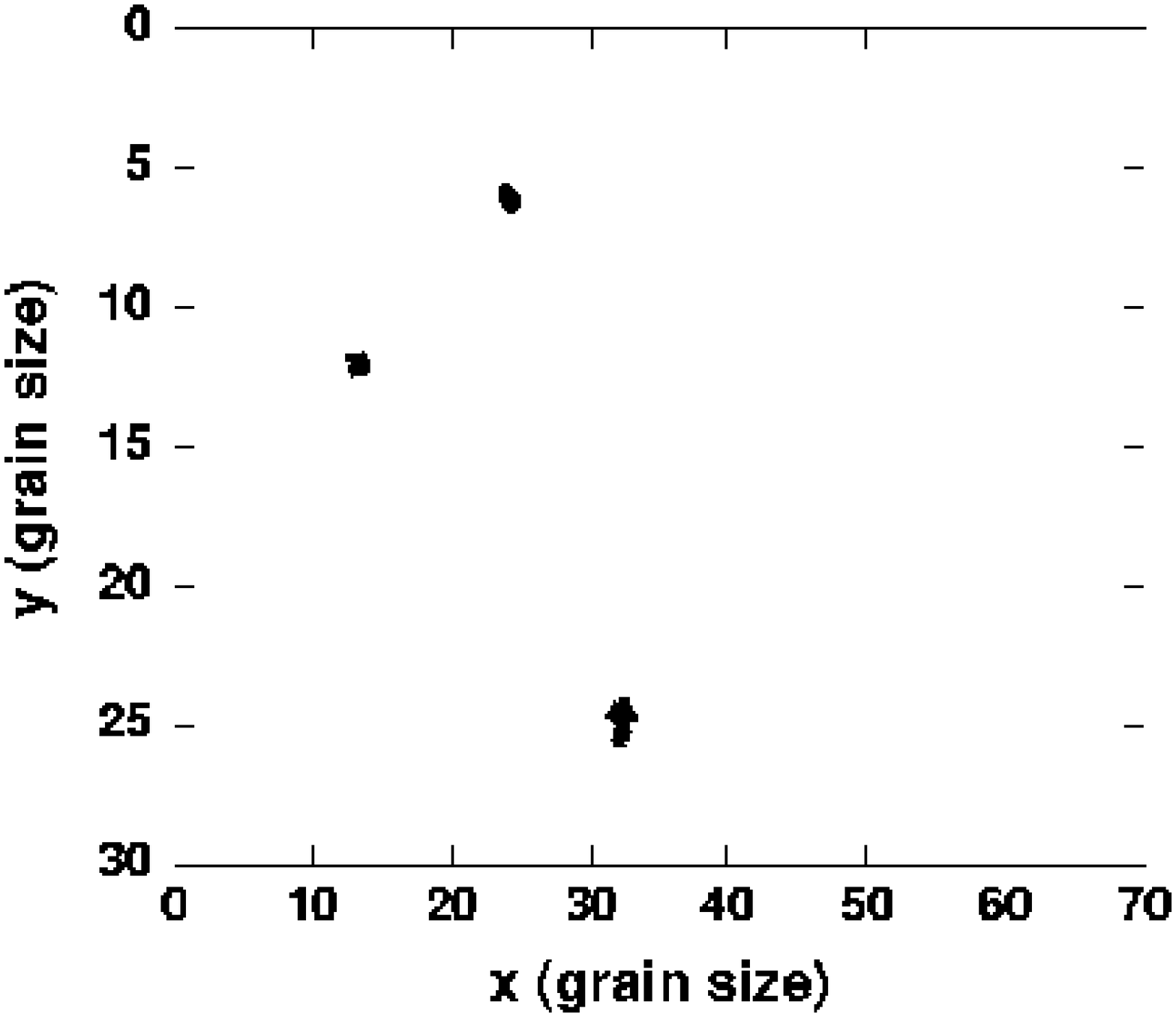}
\end{center}
\end{minipage}
\hfill\begin{minipage}[b]{0.5\linewidth}
\begin{center}
\epsfxsize=\linewidth
\epsfbox{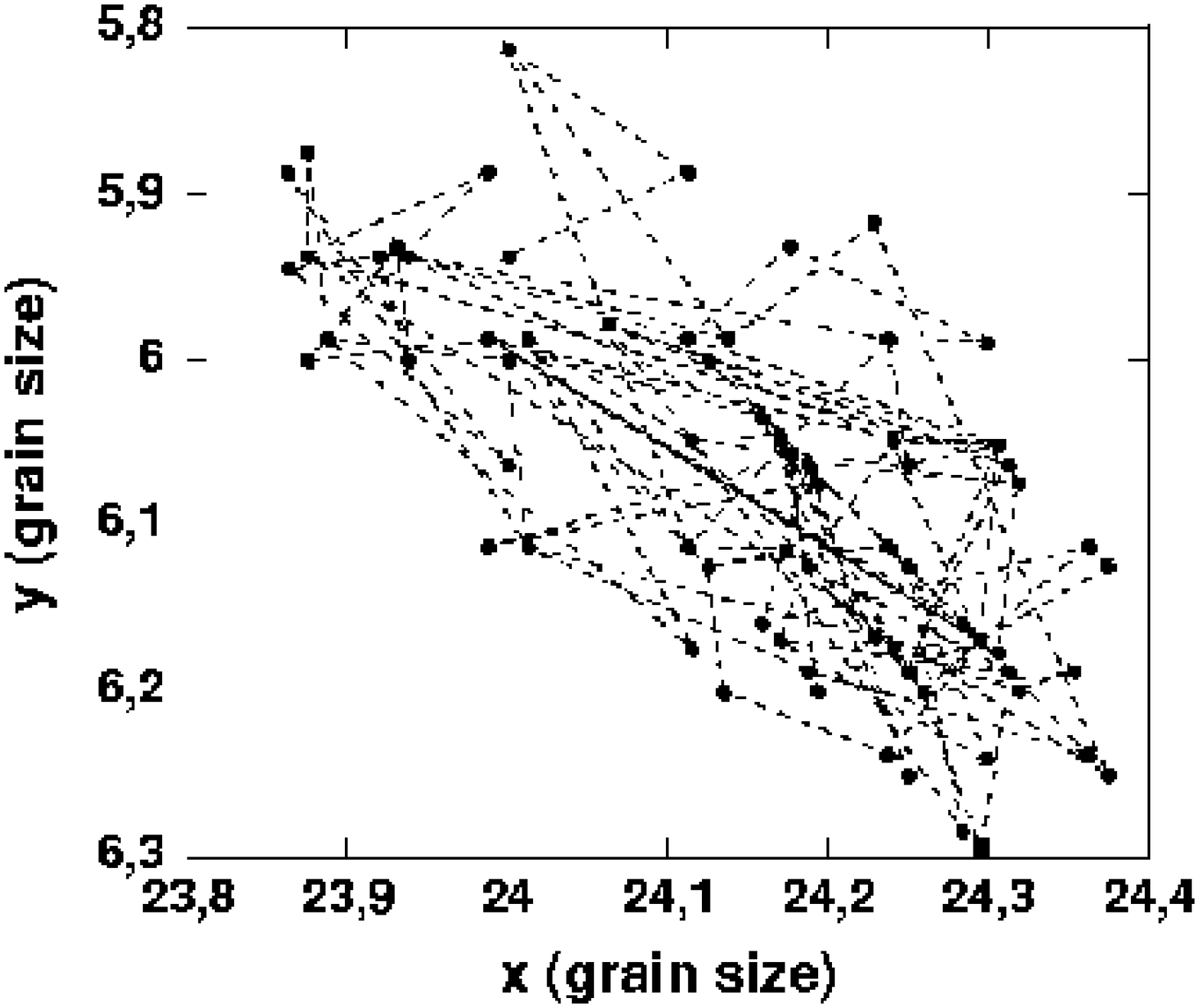}
\end{center}
\end{minipage}
\caption{Displacement of three tracer particles for experiment 4, having the
  same initial configuration and tapping amplitude as experiment 3, but with a
  lid on top. Left: Trajectory of three tracer particles. Right:
  Trajectory of one of the tracer particles (the upper particle) in detail.
\label{fig:supressed}}
\end{figure}

\begin{figure}[hh]
\begin{center}
\epsfxsize=0.6\linewidth
\epsfbox{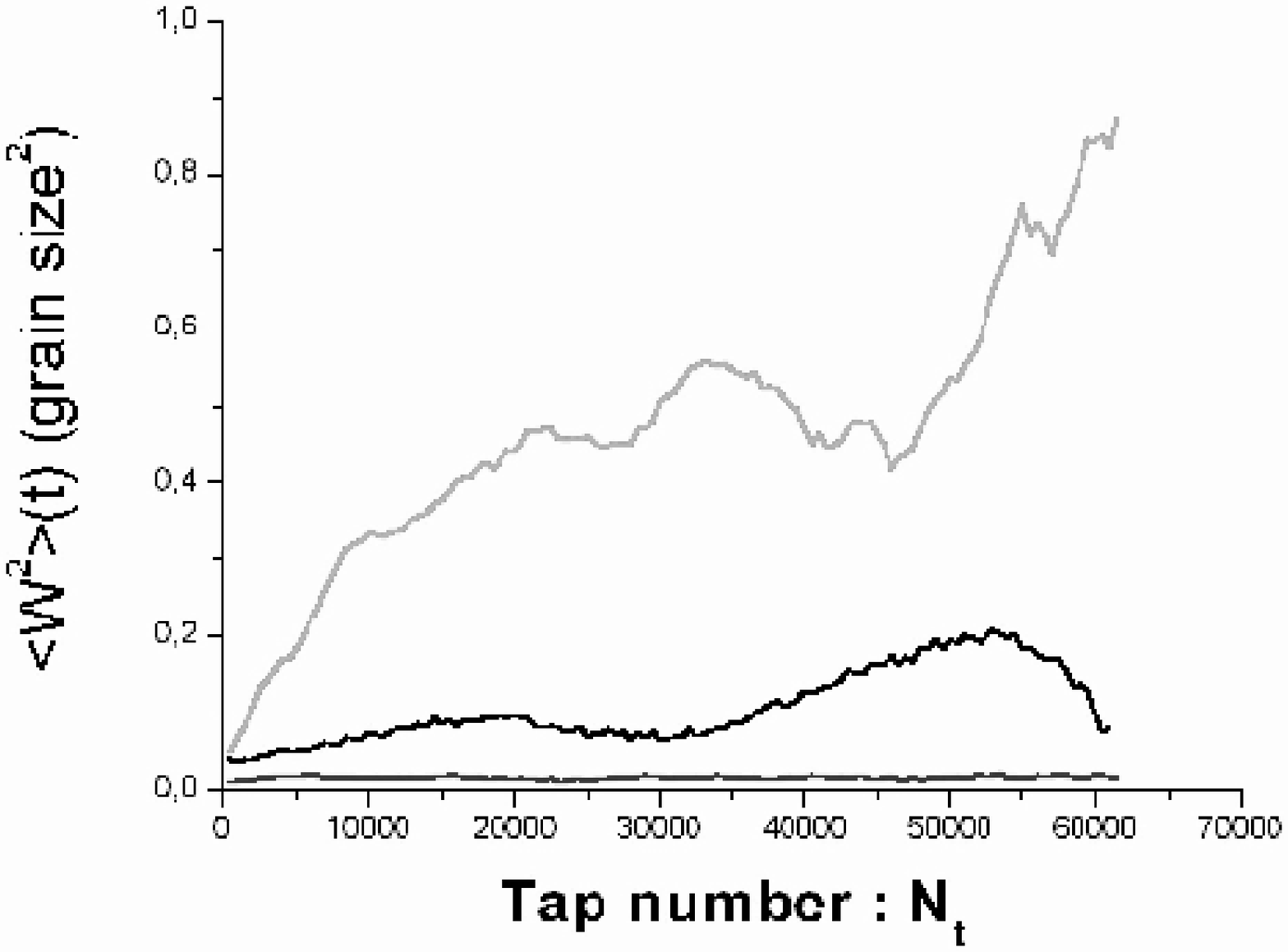}
\end{center}
\caption{$\sqrt{ \langle W^2 \rangle}$ 
as a function of tap
  number for the three tracers of experiment 4. The two curves with the lower
  displacement correspond to the two upper tracers whereas the curve with the highest
  displacement corresponds to the bottom tracer.}
\label{fig:diffusion}
\end{figure}

 The results
obtained are indeed significantly different from those with a free surface.
A direct visualization of the packing witnesses a rather strong agitation of
the grains: one can note rather pronounced collective motions of the grains
where large assemblies of particles seem to oscillate very slowly. A closer 
look at the trajectories (figure \ref{fig:supressed} left) shows however 
that these trajectories are clearly localized. Note that all displacements
shown on this graph are less than one particle diameter, even for very long
times. Note that the bottom tracer is close to the bottom plate which explains
its slightly stronger agitation. Figure \ref{fig:supressed} (right) shows the
trajectory of the upper tracer in detail and one can conclude once more that
the displacement of the grains are very small. From figure \ref{fig:diffusion}
that shows $\sqrt{\langle W^2 \rangle}$ one can conclude that once again we do
not 
observed any significant self diffusion of the
particles but that in this case the displacement is localized. Note that the
strongest displacement is seen for the bottom tracer as explained above.  

Therefore, when the convection is suppressed, we still observe
modes of motion of the packing that can lead to reorganization and compaction
but it is clear now that those modes are collective and are not associated to
self-diffusion properties of a grain.

\subsection{Outlook}
\label{sec:outlook}

In the previous sections we have shown that when applying strong tapping
($\gamma_{peak}/g>1$) to a 2D granular assembly, strong convection takes
place. This convection can be suppressed by putting a lid on top of the
granular assembly, but then the position of the lid has to be finely tuned. A
lid too close to the packing may suppress drastically the granular motion and
for a lid too far away may start convection again. Clearly, this is not a method
suited to study a problem where compaction i.e. the packing height, may vary.

Thus, we suggest to study a different system, that might allow to suppress
convection in a more adequate way and might thus be better adapted to study
the dynamics and the transport properties of an agitated packing of grains. To
do so, we have built a set-up where a large number of individually controlled
pistons are moving up and down at the bottom of a cell identical to the one
described in section \ref{sec:2dsetup}. The individual control of the pistons
allows to go from spatially and temporally coherent to more complex
agitations. Furthermore, lower tapping amplitudes can be applied such that
$\gamma_{peak}/g<1$ which will also suppress convection. We plan to measure
the displacement of one or several tracers as well as the collective modes of
reorganization, and to measure the mobility of a tracer when a given constant
force is applied. A way to do this is to use a denser particle or design a way
to pull the tracer. Further more we plan to measure the fluctuations of the
free surface. The results will then be
compared to the results obtained in 3D that will be described in the following.

\section{Compacting a 3D granular assembly}

\label{sec:3D}

The experimental set-up in 3D is set to create in the bulk a random
agitation of the grains at low level of energy and in conditions where
convection rolls due the uplift of the grains are completely suppressed.
Following the results of the 2D model system presented in the previous
section, we seek to produce an influx of energy creating a quasi randomly
agitated surface in connection with the bulk of the packing. This is the
closest we could think of a ''thermal bath''. The practical method is
described in the next subsection. Then, we expose some preliminary results
showing that the method is encouraging and might well be suited to study
compaction and jamming processes in granular assemblies and also leads to a
path where the notion of effective temperature can be precisely addressed.

\subsection{Experimental set-up}

\label{sec:set-up}

\begin{figure}[htb]
\begin{center}
\epsfxsize=0.8\linewidth
\epsfbox{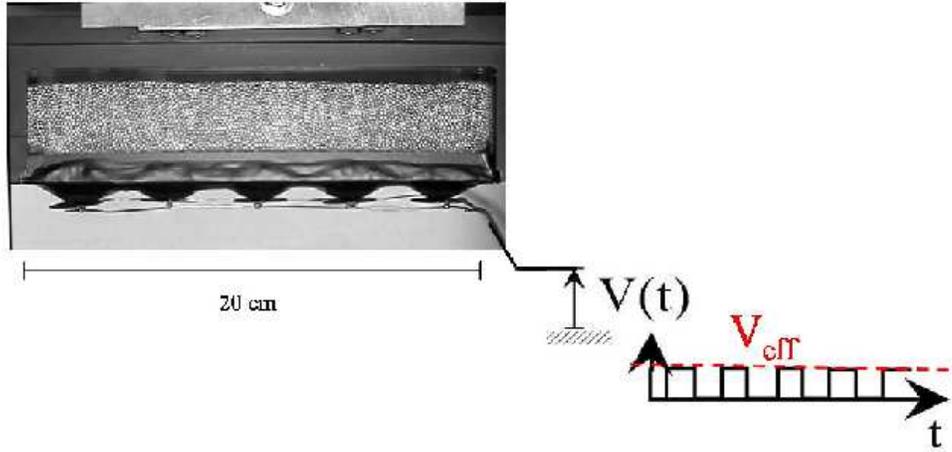}
\caption{Experimental set up: five piezoelectric transducers
  constitute the bottom of a rectangular glass box which is filled
  with $1.5 mm$ glass beads. The piezos are excited by a $380Hz$
  square signal with a maximum effective voltage $V_{eff}=35V$
\label{fig:setup}}
\end{center}
\end{figure}

The vibration device is made of five piezoelectric transducers commercially
sold as a part of a tweeter. As shown in figure \ref{fig:setup}, the transducers
are at the bottom of a conical shape paper container which is filled with $%
1.5mm$ glass beads. The surface of the granular packing confined inside the
tweeters is fixed to the observation cell and constitutes its bottom. The
cell is a hollow rectangular box of $20cm$ length, $2.3cm$ width and
around $4cm$ high. The front and rear boundaries are made of glass. The
piezos are excited by a $380Hz$ square signal with a maximum effective
voltage $V_{eff}=35V$ (figure \ref{fig:setup}). The resonance frequency of each
piezo is $f_{0}=1200HZ$. Note that the power that each piezo dissipates is
enough to make a single and lonely grain on the piezo fly up to $5mm$ high.
However, when the box is filled with grains (around $350g$), the grain
motion induced by the piezos is so small that it is not perceptible to the
eye.

\subsection{Measurement techniques}

\label{sec:measures}

\begin{figure}[hh]
\begin{center}
\epsfxsize=0.8\linewidth
\epsfbox{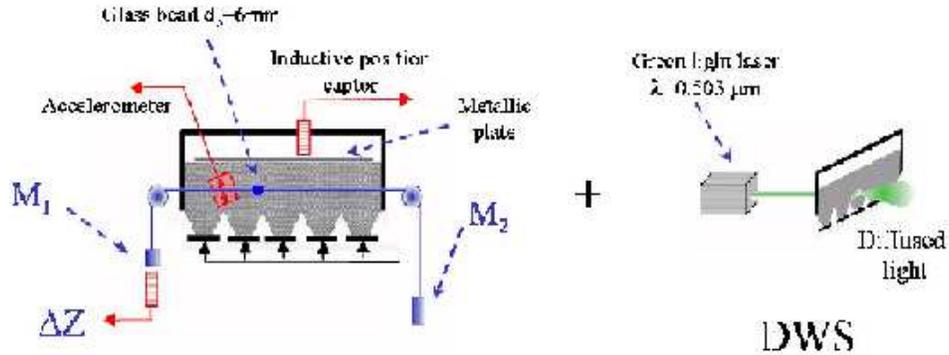}
\caption{We can measure in the system: the compaction or potential
  energy, the effective injected power, the characteristic microscopic time 
(DWS technique) and the mobility of a grain.\label{measure}}
\end{center}
\end{figure}

The cell is instrumented to measure several things (see figure
\ref{measure}) : (i) an inductive probe provides the position of a
thin metallic plate 
lying on the top granular surface and gives information about the mean 
\textbf{compaction or the potential energy} of the packing, (ii) an
accelerometer buried in the granular packing provides informations on the 
\textbf{effective injected power} (iii) we use a Diffusing Wave
Spectroscopy (DWS) technique to measure the \textbf{characteristic times
involved in the microscopic dynamics} \cite{weitz}, (iv) finally, the cell
is also designed to evaluate the \textbf{mobility} of an intruding grain.
This last point will be described more precisely in the last subchapter.

The basic idea of DWS is to make a coherent light (laser) pass through a
scattering sample in which each photon is elastically scattered multiple
times in such a way that it performs a random walk. Under this
circumstances, the light diffuses through the sample and forms an
interference pattern after it. The microscopic dynamics of the grains are
related to the dynamics of the light interference pattern. Here, a
multispeckle technique is used that was developed earlier for
colloids. The system dynamics assessment is made through the computation
of the light intensity autocorrelation function:
\begin{equation}
g_{2}(t_{w},t_{w}+t)\equiv\frac{\langle I(X,Y,t_{w})I(X,Y,t_{w}+t)
\rangle_{X,Y}}{\langle I(t_{w})\rangle_{X,Y}\langle I(t_{w}+t)\rangle_{X,Y}}-1
\label{eq:g2}
\end{equation}
where $t_{w}$ is the time of the reference image, $t$ is the time relative
to $t_{w}$ and $(X,Y)$ are the spatial coordinates on the speckle image.
The theory of light diffusion shows that under the regime of
multiple scattering (where the photons perform a random walk) it is
possible to quantitatively relate the characteristic correlation time
of $g_{2}$, $\tau$, to the mean displacement of the grains (scatterers)
\cite{weitz}. However, in our case, preliminary measurements show that
due to the relatively small number of grains across the cell (about 14 grains), the
condition of a fully developed random walk for the light path is
probably not fully achieved and thus, the possibility to trace back
quantitatively the average motion of the grains is somehow
complicated. On the other hand, at this stage, the correlation
function is still a good and sensitive indicator for the granular
dynamics.

To prepare the system, we poured the grains into the box and gave gentle
taps to compact the grains such as to ``erase'' the structural memory of the
pouring and to prepare the system at a given packing fraction. Next, we
placed the thin metal lid on the surface and let the laser go through the
packing. We left it go for one hour in order to stabilize the laser as well
as the initial stress relaxations in the sample. Note that in these
experiments we did not put the intruding grain because it would have
interfered with the other measurements. Once stabilized, we turned on the
piezos at their maximum capacity ($35V_{eff}$, which corresponds to a r.m.s.
acceleration measured by the accelerometer of $\langle\gamma^{2}\rangle^{1/2}=0.24m/s^{2}<<g$). After an hour of vibration we changed the
voltage value. We continued the vibration changing the voltage each hour,
measuring the position of the surface and 
recording the speckle images during each
of these one-hour vibration intervals. Actually, we made two different
experiments (labeled experiment 1 and experiment 2) with different voltage
variation sequences.

\begin{figure}[h]
\begin{minipage}[b]{0.5\linewidth}
\begin{center}
\epsfxsize=\linewidth
\epsfbox{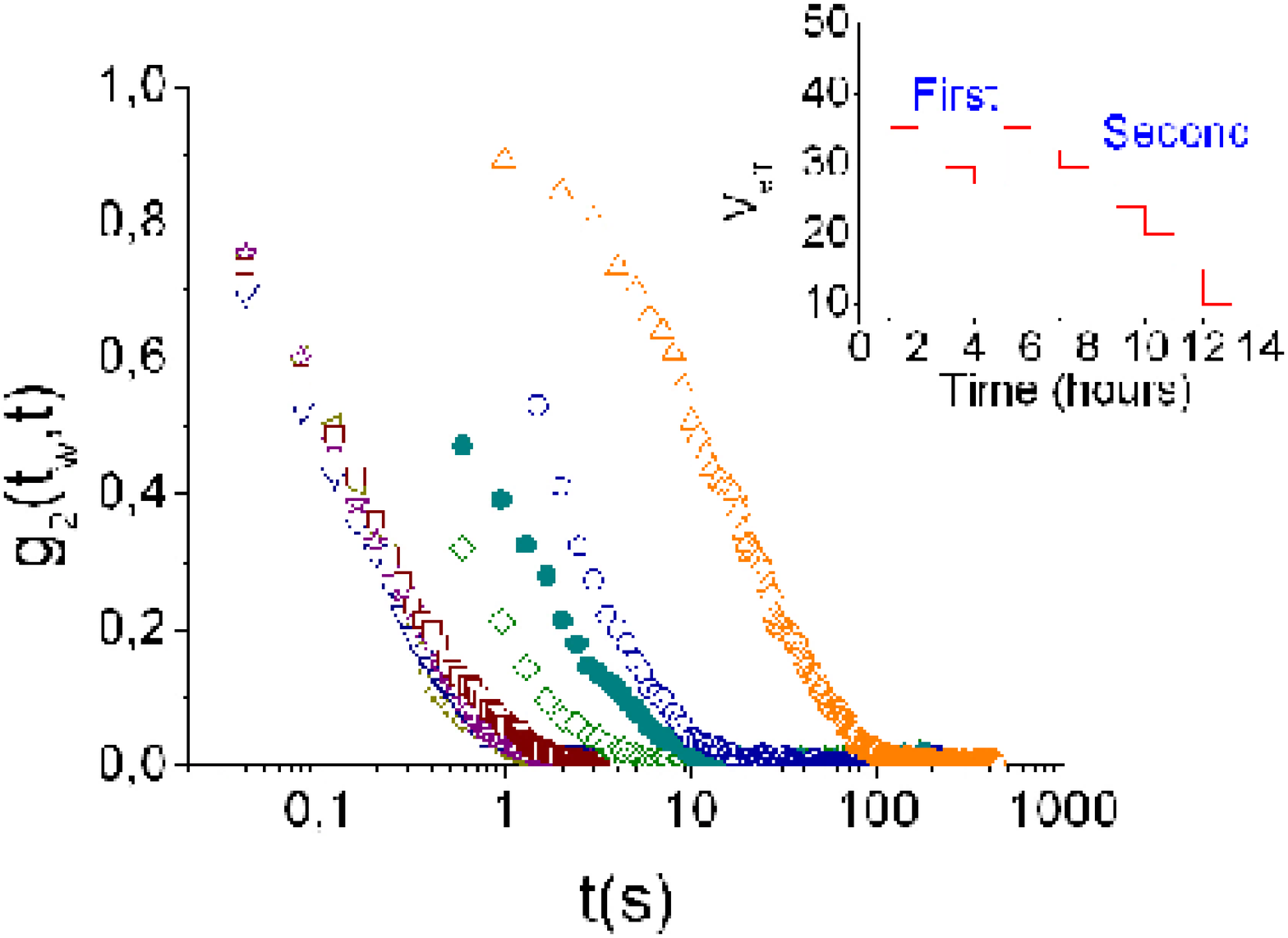}
\end{center}
\end{minipage}
\hfill\begin{minipage}[b]{0.5\linewidth}
\begin{center}
\epsfxsize=\linewidth
\epsfbox{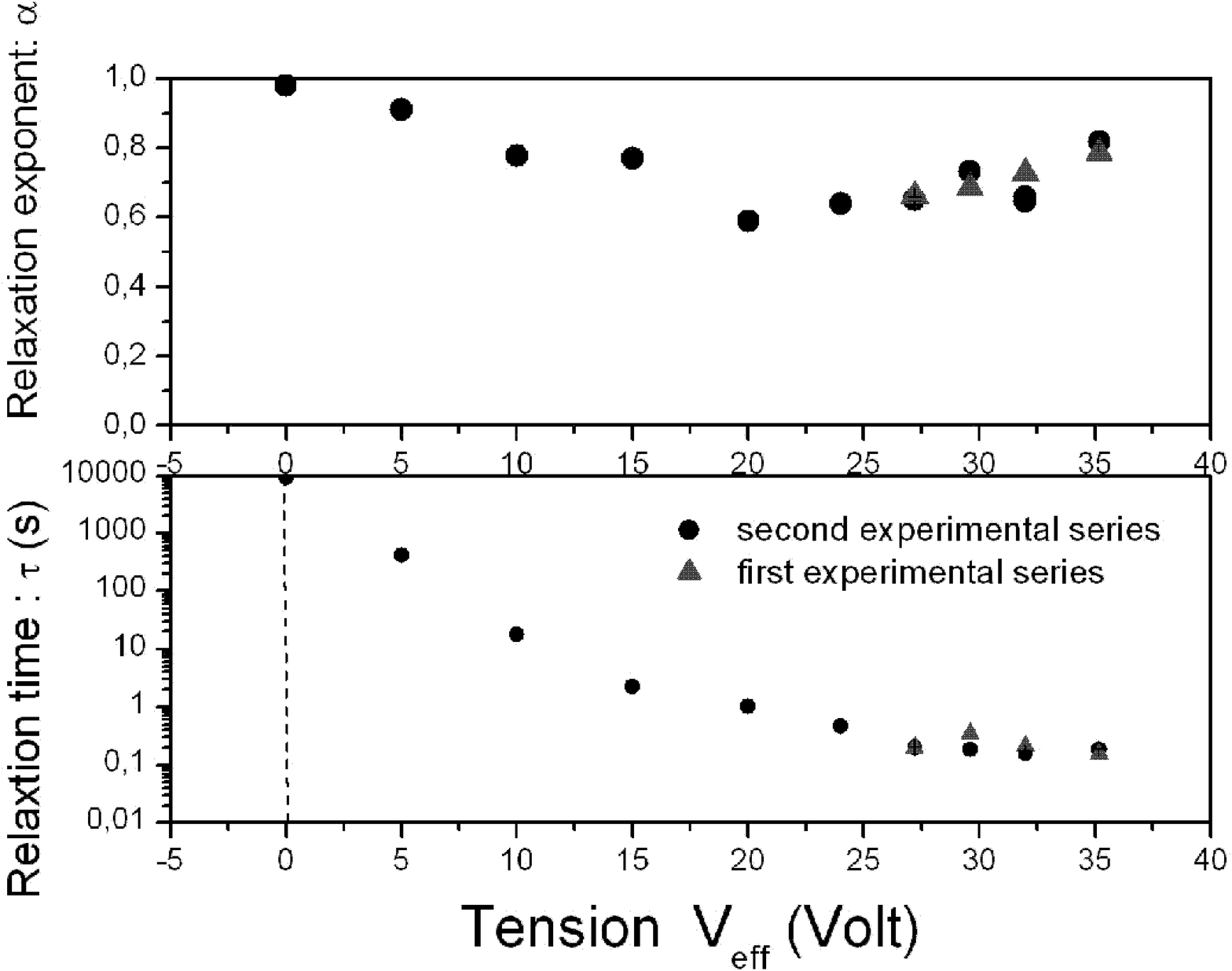}
\end{center}
\end{minipage}
\caption{Left: Correlation functions for the second descent of experiment 1: 
($\triangleleft$)$V_{eff}=35V$, ($\triangledown$)$V_{eff}=32V$, ($\bigstar$)
  $V_{eff}=29.6V$, 
($\square$)$V_{eff}=27.2V$, ($\lozenge$)$V_{eff}=24V$, 
($\bullet$)$V_{eff}=20V$, ($\circ$)$V_{eff}=15V$, 
($\vartriangle$)$V_{eff}=10V$. Inner frame: Voltage variations in 
experiment 1: we began with a first descent from 35 to $27.2V$; 
then, we came back to $V_{eff}=35V$ to make a second descent down to
$V_{eff}=10V$. Right: Behavior of $\alpha$ and $\tau$, parameters of the
  fitting stretched exponential to the correlation function $g_2(t,t_w)$, as a
  function of effective voltage for the first and second voltage
  descents of experiment 1.
\label{fig:g2}}
\end{figure}

\subsection{Preliminary results}

\subsubsection{Experiment 1 - compaction and microscopic dynamics}
\label{sec:exp1}

The inner frame of figure \ref{fig:g2} (left) shows the voltage variations
that we followed in this experiment. We began with a \textit{first descent}
from $V_{eff}=35V$ to $V_{eff}=27.2V$; then, we came back to $V_{eff}=35V$
and made a \textit{second descent} down to $V_{eff}=10V$. We did this with
the idea of reaching a stationary state during the first strong vibration
steps and to compare the measurements for the same voltage values in the
second voltage descent. The correlation functions $g_{2}(t_{w},t)$ of figure 
\ref{fig:g2} (left) fit well with a stretched exponential
$y(t)=exp(-(t/\tau)^{\alpha})$.

\begin{figure}[h]
\begin{minipage}[b]{0.5\linewidth}
\begin{center}
\epsfxsize=\linewidth
\epsfbox{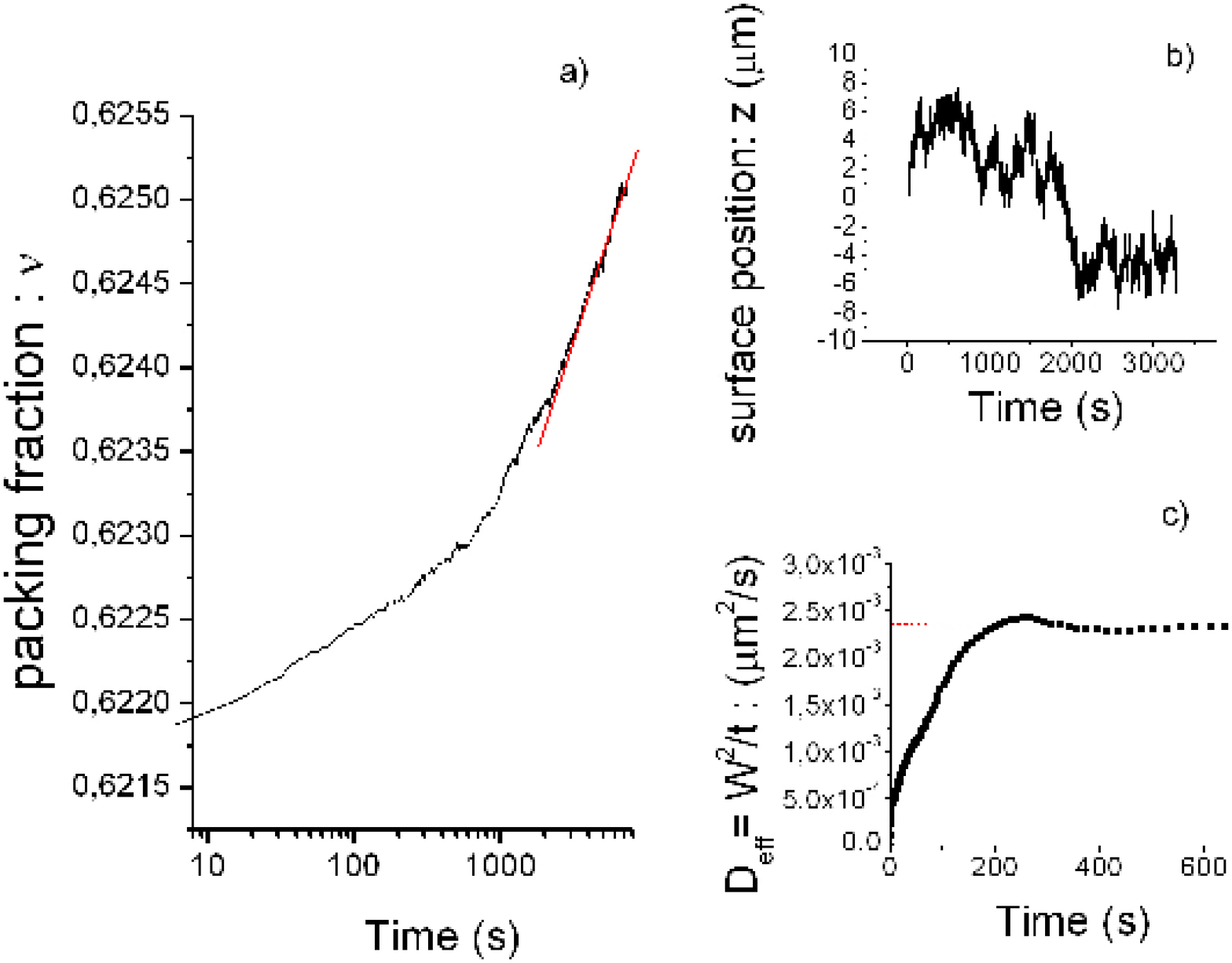}
\end{center}
\end{minipage}
\hfill\begin{minipage}[b]{0.5\linewidth}
\begin{center}
\epsfxsize=\linewidth
\epsfbox{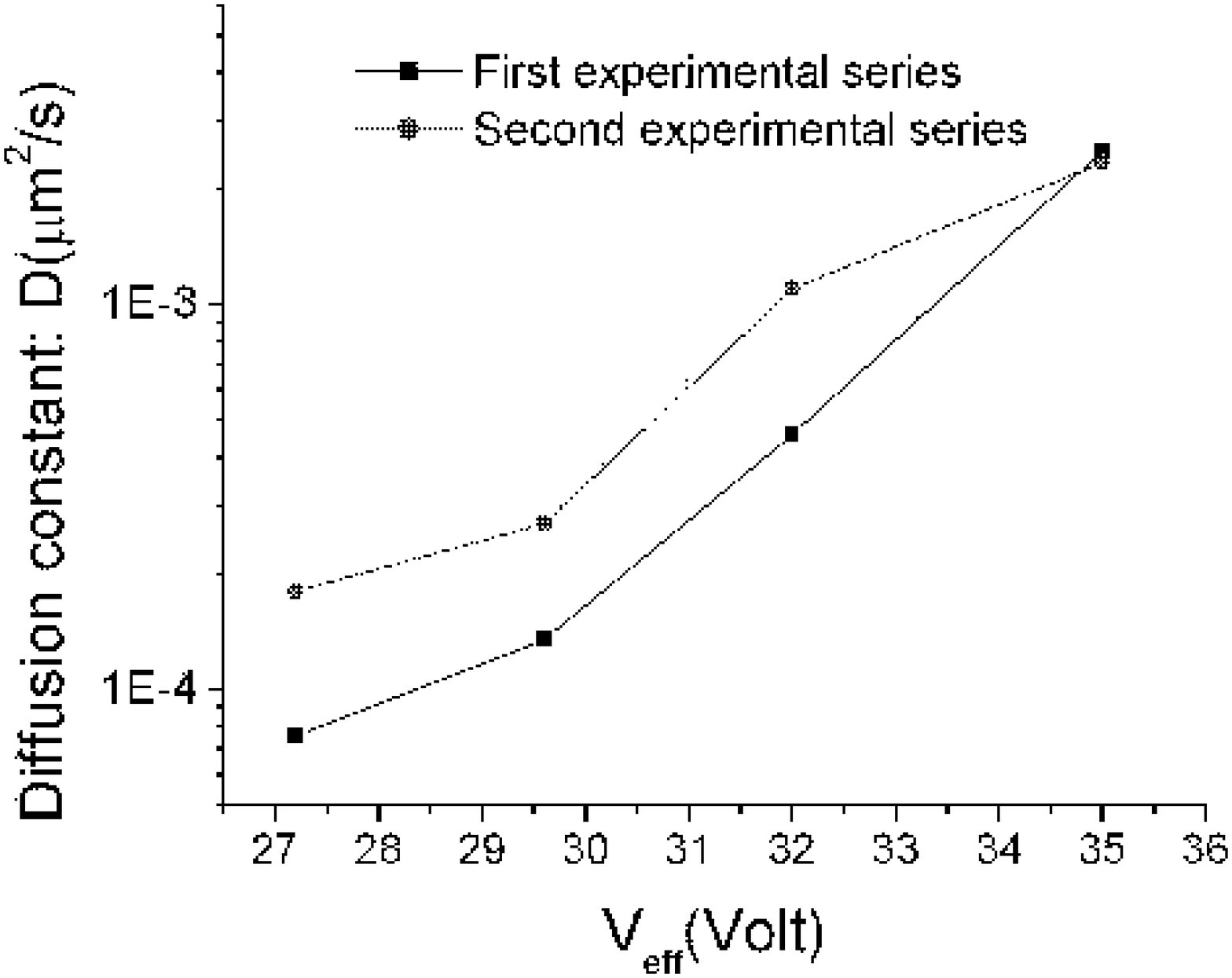}
\end{center}
\end{minipage}
\caption{Left: a) Initial compaction at $V_{eff}=35V$. A steady state is not
  observed. b) Short time diffusion-like fluctuations of surface's
  position during vibration. It corresponds to the second descent at
  $V_{eff}=35V$. c) Surface's position fluctuates following a simple
  diffusion; the diffusion constant is calculated from position's
  roughness. Right: Effective diffusion of the height of the surface vs
  effective voltage}
\label{surface}
\end{figure}

On the right of figure \ref{fig:g2} we can see the behavior of $\alpha$
and $\tau$ as a function of voltage. From the comparison of the first and
second descents we don't observe any ageing effect, since the data
corresponding to both of them are similar. Noteworthy, for the value of $%
\alpha$ other authors \cite{kabla} have reported a constant value of $0.8$
for similar systems.

From the position of the surface we can, in principle, calculate the packing
fraction $\nu$ and the potential energy $E$ of the system. However,
despite the high precision data that we have for the surface position, we
can only roughly estimate the corresponding \emph{absolute} value of the
packing fraction ($5\%$ error). This is due to the lack of precision in
determining the total volume of the cell+tweeters. The left of figure
\ref{surface} (a) shows the compaction dynamics of the system during
the first hour of vibration. After an initially ''fast'' decrease of
the height, we
indeed observe a slow logarithmic-like compaction process. Note that a
stationary state is not yet reached. Plot (b) of the left of figure \ref{surface} shows
a typical measure of the position of the surface during vibration at $V_{eff}=35V$.
It corresponds to the second descent, i.e. five hours after the beginning of
the vibration and though it seems that the system had reached a stationary
state, there is still a slow systematic compaction not perceptible at
such time scale. Interestingly, the position of the surface seems to undergo wild
fluctuations expanding and compacting in an irregular manner. However, the
spatial displacement of the level of the surface is of the order of microns,
much less than a grain size. So here we are talking about a collective
macroscopic behavior since individual grain migration is not possible. This
surface dynamics should be associated with the typical size of grain
rearrangements in its surrounding (slipping of contacts and rotations)
performed more or less randomly per unit time and integrated over the cell
height. This is why we now investigate the dynamics of the
fluctuations more precisely.

From the signal of figure \ref{surface} (b) we calculated the roughness $%
W=\langle(z-\langle z\rangle)^{2}\rangle^{(1/2)}$, where $z$ is surface's
position, as a function of time and we found that the surface fluctuates as
a simple diffusion process since $D_{eff}=W^{2}/t$ tends to a constant, the
effective diffusion constant, for long times (figure \ref{surface},
c)). In
figure \ref{surface} right we show this constant for different values of the
effective voltage for both experimental series. Now it is clear that there
is a difference between the first and the second descent which could be a
signature of the ageing of the packing. However a final conclusion is not
possible here since the data corresponds to a single realization.

\begin{figure}[hh]
\begin{center}
\epsfsize=0.7\linewidth
\epsfbox{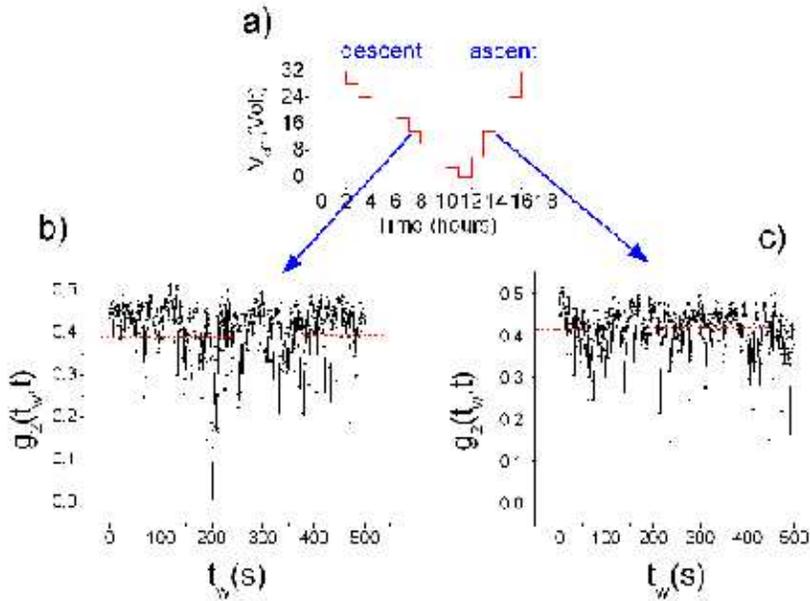}
\caption{a) Voltage variations in experiment 2.
 b) Autocorrelation intensity function (eq. \ref{eq:g2}) for variable
     $t_w$ and constant $t=0.5s$ at $V_{eff}=14 V$ of the 
     descending part of voltage variation. c) The same as
     (b) but for the ascending part. The correlation drops off from its mean
     value as a result of the intermittent dynamic of the system near
     jamming transition.\label{exp2}}  
\end{center}
\end{figure}

\subsubsection{Experiment 2 - Dynamical intermittency}

\label{sec:exp2}

\begin{figure}[hh]
\begin{center}
\epsfsize=0.5\linewidth
\epsfbox{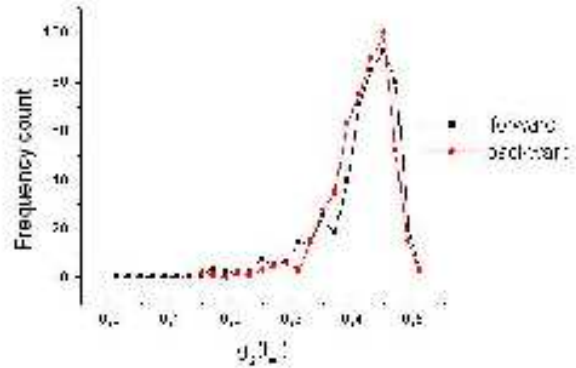}
\caption{Histogram of the correlation curves of figure \ref{exp2}. The
  deviation from a Gaussian curve results from intermittency
  \cite{Lucas}. From comparison of the forward and backward data we
  don't find evidence of ageing for this particular case.\label{hystCI}} 
\end{center} 
\end{figure}

One of the first things we noticed when we began to work with vibrated
systems was that the speckle dynamics during vibration was not continously
varying but would rather show intermittent or irregular behavior. For a
given vibration intensity we observed that the speckle changed with a
typical characteristic decorrelation time, but the value of
this time would often present big variations depending on which initial
speckle pattern was chosen. Basically there were phases where the
decorrelation would be regular and steady but others where an intense
dynamics would drastically accelerate the speckle decorrelation.
Interestingly, this kind of behavior was observed recently in thermal
systems near jamming transition by Cipelletti {\it et al.} \cite{Lucas}. Moreover,
these authors have proposed a way to quantitatively characterize the
intermittency using the autocorrelation intensity function (eq. \ref{eq:g2})
obtained from the DWS technique. The idea is to keep fixed the time interval
between correlating images $t$ and to vary the reference image $t_{w}$.
If the speckle dynamics were continuous this analysis would result in a kind
of noise around a constant correlation value depending on $t$. However, huge
``catastrophic'' events would give low correlation values, highly deviated
from the mean. The statistical analysis of such events leads to a
quantitative characterization of the systems dynamics \cite{Lucas}. Following
these ideas we made an experiment similar to the previous one but with
different voltage variations (fig. \ref{exp2} (a)). In (b) and (c) of figure 
\ref{exp2} we observe the autocorrelation intensity function for
variable $t_{w}$ and constant $t=0.5s$ for the same vibration
intensity ($V_{eff}=14V$), (b) corresponding to the descending part of
the voltage
variation and (c) to the ascending part. It is possible to appreciate
how the correlation
frequently drops off from its mean value. In figure \ref{hystCI} is shown
the histogram of the data from plots of figure \ref{exp2}. If there was an
ageing effect the two histograms should be different, which is not the case.
However, the deviation of the histograms from a Gaussian one to the low
values of $g_{2}(t,t_{w})$ is an evidence of intermittency
\cite{Lucas}.

\subsubsection{Mobility of an intruding grain}

\label{sec:mobility}

\begin{figure}[hh]
\begin{center}
\epsfsize=0.5\linewidth
\epsfbox{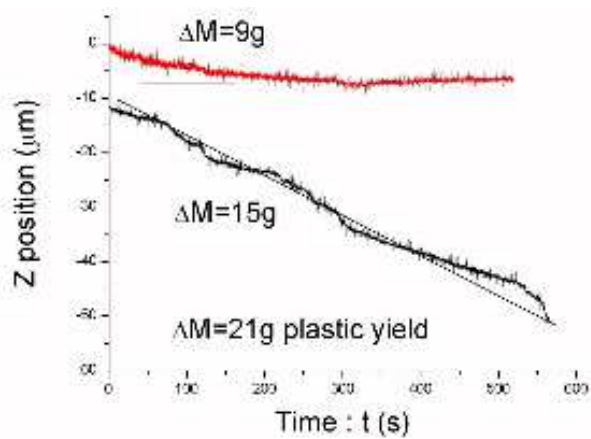}
\caption{Measurement of the displacement of the intruding grain during
  vibration for different $\Delta M$.\label{mobility}}
\end{center}
\end{figure}

In this experiment we tried to measure the mobility of a grain inside the
vibrated granular system by forcing an intruding grain to move trough it. We
glued the intruding grain to a fishing thread which passes through the
granular assembly and that is tightened by two masses of different weight in
an Atwood's machine-like configuration (see figure \ref{measure}). The
intruder's position is monitored by an inductive captor that measures the
displacement of one of the tightening masses. The size of the intruding
grain is of $6mm$ in diameter. Figure \ref{mobility} shows three different
regimes of the mobility as a function of $\Delta M=M_{2}-M_{1}$ (see figure 
\ref{measure}). For $\Delta M=9g$ the intruding grain is initially moving but
suddenly gets jammed. A continuous movement around a constant velocity is
observed for $\Delta M=15g$. Finally, for $\Delta M$ around $25g$ there is a
plastic yield and the intruding grain moves without resistance through the
granular material. Although these are only preliminary results they show that
this technique allows for interesting mobility
measurements.

\section{Conclusions}

We have shown the results of two experiments of vibration of a granular
material, one in 2D and one in 3D, where both macroscopic and microscopic
features of the flow can be measured and eventually related. In the case of
the two dimensional assembly with free surface, it was also clear that the
long time dynamics is dominated by convection and that dispersion due to
local self diffusion is not a dominant feature. While moving into the huge
convection rolls, the grains would roughly conserve their neighbors for a
very long time. We could not measure any self-diffusion characteristics for
the whole durations of the experiments and this conclusion still holds even
when convection (but not the granular agitation) is suppressed by imposing a
lid on the top. Consequently, from these direct observations in 2D, we came
to the conclusion that in all standard shaking experiments, the observed
compaction steady states could result from a competition between compaction
due to vibration and expansion due to convection. Moreover, we stress on two
important points.  First, the suppression of convection is quite important
to a controlled study of compaction dynamics and to access the notion of
``effective temperature''more easily. Second, in the very dense
phase, the notion of self-diffusion of a particle is unlikely to be a
relevant parameter suited to characterize the jammed dynamics. Basically,
the motions of the grains are localized in the cage formed by their
neighbors and their jumps statistics within the cage reflect
\emph{collective} 
modes of reorganization. These dynamical processes are indeed
sufficient to create global compaction/decompaction processes without motion
of a grain out of its neighbor's cage.

We managed to achieve a convection free vibration in the three dimensional
experiment by putting the grains in direct contact with five piezoelectric
transducers. Then, we were able to study the compaction dynamics in the weak
vibration limit, i.e. imposed by  accelerations much smaller than gravity.
A study of compaction under this circumstances is quite new. We observed a
very slow density relaxation and diffusion of the free surface that we will be able to characterize more
properly in experiments actually in progress. Interestingly, the magnitude
of volume or density fluctuations indicates that there is no grain diffusion
during vibration, thus the  compaction and reorganization dynamics of the
bulk is solely due to collective features. Another advantage of weak
vibrations is that we managed to observe and measure intermittent dynamics
which creates a link with results on jammed colloidal phases. This becomes
important since, as pointed out by L. Cipelletti et al. \cite{Lucas},
temporal heterogeneities are a fundamental feature of the dynamics of jammed
systems. Interestingly, surface diffusion as well as intermittent dynamics were also found 
in stochastic compaction models \cite{AnitaProc,Luck}. Moreover, it has been proposed that a practical way to
experimentally measure effective temperature is to analyze intermittent
events \cite{ritort}. These preliminary results are encouraging. We think
that we will be able to extract very interesting information from this
system once we had done all the calibrations needed and improved our
procedures. Also, a central point is to manage to reach the stationary
state, which seems not easy since we work with extremely weak vibrations. In
fact, one of the things that we want to do is to increase the power
injection to the system and to randomize it as much as possible. Certainly,
it would be very interesting to couple together all the different measures
that we can make on the system. With this, we could try to relate the
microscopic features (DWS and mobility measurements) to macroscopic ones
(compaction dynamics and density fluctuations). Of course, the next obvious
step is to obtain the measurements of density and its fluctuations (or,
equivalently, energy and its fluctuations) to compare them to the numerical
results of A. Fierro et al. \cite{naples}.
\vspace{5pt}

{\bf Acknowledgments} This project is part of ECOS M03P01 and GC is supported 
by CONACYT and DGEP.


\begin{thebibliography}{99}

\bibitem{LiuNature} A. J. Liu and S. R. Nagel, Nature {\bf 396},
  21 (1998).

\bibitem{DaCruz02} F. da Cruz, F. Chevoir, D. Bonn and P. Coussot,
  Phys. Rev. E {\bf 66}, 051305 (2002).

\bibitem{Edwards} S. F. Edwards, Physica A {\bf 249}, 226-231 (1998)

\bibitem{Kurchan} A. Barrat, J. Kurchan, V. Loreto and M. Sellitto,
  Phys. Rev. Lett. {\bf 85}, 5034 (2000); Phys. Rev. E, {\bf 63}, 051301
  (2001). 

\bibitem{Ono} F.K. Ono, C.S. O'Hern, D.J. Durian, S.A. Langer, A.J. Liu and
  S.R. Nagel, Phys. Rev. Lett. {\bf 89}, 095703 (2002).

\bibitem{AnitaGen} A. Metha et G.C. Barker, Phys. Rev. Lett. {\bf 67}, 394 
(1991).

\bibitem{danna} G. D'Anna, A. Barrat and V. Lorento, F. Nori, Nature {\bf 424},
  909 (2003). 

\bibitem{Coniglio} A. Coniglio and M. Nicodemi, Physica A {\bf 296},
  451 (2001). 


\bibitem{NowakPRE} E.R. Nowak, J.B. Knight, E. Ben-Naim, H.M. Jaeger and
  S.R. Nagel, Phys. Rev. E {\bf 57} 1971 (1998).

\bibitem{Pouliquen} O. Pouliquen, M. Belzons and M. Nicolas, cond-mat/0305659.

\bibitem{naples} A. Fierro, M. Nicodemi and A. Coniglio,
  Europhys. Lett. {\bf 59}, 642-647 (2002).

\bibitem{ClementRev} E.Cl\'{e}ment, ''Granular packing under vibrations '', in
    \textit{ '' Physics of dry granular media''} p. 585, Ed. J.P. Hovi,
    H. Herrmann and S. Luding (Kluwer Acad. Publisher, 1998).

\bibitem{Evesque} P. Evesque et J. Rajchenbach, Phys. Rev. Lett. {\bf 61}, 44
  (1989).

\bibitem{Laroche} C. Laroche, S. Douady and S. Fauve, J. Phys. (Paris) 
{\bf 50} 699 (1989). 

\bibitem{Clement92} E. Cl\'{e}ment, J. Duran and J. Rajchenbach, Phys. Rev.
Lett. {\bf 69}, 1189 (1992).


\bibitem{Ovarlez} G. Ovarlez, C. Fond, E. Cl\'ement, Phys. Rev. E {\bf 67},
  60302 (2003).


\bibitem{Duran} J. Duran, T. Mazozi, E. Cl\'{e}ment, J. Rajchenbach,
  Phys. Rev. E. {\bf 50}, 3092 (1994).

\bibitem{PhilippePRL} P. Philippe and D. Bideau, Phys. Rev. Lett. {\bf 91}
  104302 (2003).

\bibitem{PhilippePRE} P. Philippe and D. Bideau, Europhys. Lett. {\bf 60}, 677
  (2002).

\bibitem{KnightPRE} J.B. Knight, C.G. Fandrich, C.N. Lau, H.M. Jaeger and
  S.R. Nagel, Phys. Rev. E {\bf 51}, 3957 (1995).


\bibitem{KnightPRE96} J.B. Knight, E.E. Ehrichs, V.Y. Kuperman, J.K. Flint,
  H.M. Jaeger and S. Nagel, Phys. Rev. E {\bf 54}, 5726 (1996).

\bibitem{weitz} D.A. Weitz and D.J. Pine. Diffusing-wave 
spectroscopy. In Wyn Brown, editor, \textit{Dynamic light scattering: The
method and some applications}, volume 49 of 
\textit{Monographs on the physics and chemistry of material}, pages 
652-720. Oxford University Press, Oxford, 1993.

\bibitem{kabla}A. Kabla and G. Debr\'{e}geas, cond-mat/0303560.



\bibitem{Lucas}L. Cipelletti, H. Bissig, V. Trappe, P. Ballesta and S.J.,
Phys.:Condens Matter {\bf 15}, S257, (2003).

\bibitem{AnitaProc} A. Metha and J.M. Luck, {\it Shaken not stirred: why
    gravel packs better than bricks}, present proceedings.

\bibitem{Luck} J.M. Luck and A. Metha, European Journal of Physics B, {\bf 35}, 399 (2003)

\bibitem{ritort} A. Crisanti and F. Ritort, cond-mat/0307554.


\end{thebibliography}
\end{document}